\documentstyle[psfig,doublespace]{article}
\def\etal{\mbox{\it et al.\ }}
\def\Da{\mbox{Damk$\ddot{\rm{o}}$hler }}

\begin{document}

\title{Compressibility effects on the scalar mixing in reacting
homogeneous turbulence}

\author{D. Livescu and C.K. Madnia\\
Department of Mechanical and Aerospace Engineering\\ 
State University of New York at Buffalo\\ 
Buffalo, NY 14260}
\date{}
\maketitle

 \begin{abstract}

The compressibility and heat of reaction influence on the scalar mixing in 
decaying isotropic turbulence and homogeneous shear flow are examined via data
generated by direct numerical simulations (DNS). The reaction is modeled as  
one-step, exothermic, irreversible and Arrhenius type. For the shear flow
simulations, the scalar dissipation rate, as well as the time scale ratio
of mechanical to scalar dissipation, are affected by compressibility and
reaction. This effect is explained by considering the transport equation
for the normalized mixture fraction gradient variance and the relative 
orientation between the mixture fraction  gradient and the eigenvectors of the
solenoidal strain rate tensor.

 \end{abstract}

%

\section{Introduction}

Recent studies of scalar mixing in incompressible turbulence 
(e.g. Overholt \& Pope 1996; Jaberi \etal 1996; Slessor, Bond \& Dimotakis
1998; Livescu, Jaberi \& Madnia 2000) have enhanced our understanding of the 
mixing mechanism, the behavior of the scalar PDF and the evolution of scalar 
moments. However, fundamental studies of scalar mixing in compressible 
turbulence are scarce. Blaisdell, Mansour \& Reynolds (1994) show that the 
dilatational velocity has negligible contribution to the scalar flux in 
homogeneous turbulence. It is not clear, 
however, how the compressibility affects the scalar moments or the scalar 
dissipation. A better understanding of the compressibility effects is important
in the analysis of the combustion processes. The main objective of 
this work is to study the influence of compressibility and heat release on
the evolution of the scalar dissipation rate in homogeneous turbulence.

\section{Numerical methodology and DNS parameters}

In order to assess the influence of compressibility on the mixing process, 
direct numerical simulations (DNS) of decaying isotropic and homogeneous 
sheared turbulence are performed under reacting (heat - releasing) and 
nonreacting conditions. In the nonreacting cases a passive scalar is simulated
along with the turbulent field. The compressible form of continuity, momentum,
energy and species mass fractions transport equations are solved using the 
spectral collocation method. In the reacting cases, the chemical reaction is 
modeled by a one-step irreversible Arrhenius-type reaction. In order to 
assess the influence of heat of reaction on the mixing process, the statistics
pertaining to the mixture fraction $Z$ are extracted from the two reacting 
scalar fields and compared to those obtained for a passive scalar in the 
nonreacting cases. The viscosity varies with the temperature according with a 
power law and $Le=1$ in all cases. 

For the shear flow simulations, the velocity fluctuations are initialized
as a random solenoidal, three-dimensional field with Gaussian spectral density
function (with the peak at $k_{0v}=10$) and unity rms.  
The initial pressure fluctuations are evaluated from a Poisson equation 
(except for the high Mach number case for which they are set to zero). For 
the decaying isotropic simulations the initialization proposed by Ristorcelli 
\& Blaisdell (1997) is employed. The variables are then allowed  to decay 
until they develop realistic 
turbulence fluctuations before the scalars are initialized. This 
corresponds to time $t=0$ on all figures. The value of the Reynolds number
at $t=0$ is $Re_{\lambda_0}=50$. The simulations are stopped at 
$t/\tau_t=14.4$ such that the scalar variance becomes two orders of magnitude 
less than the initial value. For the shear flow cases $Re_{\lambda_0}=21$ 
initially, and increases to values between $93-110$ in the direction of the 
mean velocity at the end of the simulations. The shear flow simulations 
are stopped between $18<St<20$, such that the integral scales remain small 
compared to box size and the Kolmogorov microscale is larger than the grid 
size. This corresponds to $t/\tau_t$ in the range  $10.4$ to $11.5$.

The scalar fields are initialized as ``random blobs'', with double-delta PDFs 
(Overholt \& Pope, 1996). The initial length-scale of the scalar field
is controlled by changing the location, $k_{0s}$, of the peak of the Gaussian 
spectrum used to generate the scalar field. The variables are time advanced in
physical space using a second order accurate Adams-Bashforth scheme. The code 
developed for this work is based on a fully parallel algorithm and uses the 
standard Message Passing Interface (MPI).

In order to examine the compressibility effects on the scalar mixing, cases 
with different initial value of the turbulent Mach number, $M_{t_0}$, are
considered. The heat release effects are investigated by varying the values  
of the heat release parameter, $Ce$, computational \Da number, $Da$, and the 
initial lengthscale of the scalar field. All reacting cases have the value of 
the Zeldovich number, $Ze=8$. Table \ref{tab1}  summarizes the cases 
considered for this study. The isotropic turbulence cases are labeled with $i$
and the shear flow cases with $s$. For the shear flow cases the value of the 
initial nondimensional mean shear rate, $S_0^*=\frac{S(2K)}{\epsilon}$, where 
$K$ is the turbulent kinetic energy and $\epsilon$ is the viscous dissipation,
is in the range dominated by nonlinear effects. 
$S=\partial\tilde{u_1}/\partial x_2$  ($\tilde{\mbox{ }}$ is the Favre 
average) is the mean shear rate.  

\begin{table}[h]
\caption{Parameters for the DNS cases.}
\begin{center}
\begin{tabular}{cccccc}
\hline
\it Case $\#$ &\it $M_{t_0}$ &\it $S_0^*$ &\it $Da$ &\it $Ce$ &\it $k_{0_s}$\\
\hline
$i1$ & 0.2 & 0 & 0    & 0    & 4\\
$i2$ & 0.35& 0 & 0    & 0    & 4\\
$i3$ & 0.5 & 0 & 0    & 0    & 4\\
\hline
$s1$ & 0.1 & 7.24 & 0 & 0 & 4\\
$s2$ & 0.2 & 7.24 & 0 & 0 & 4\\
$s3$ & 0.3 & 7.24 & 0 & 0 & 4\\
$s4$ & 0.4 & 7.24 & 0 & 0 & 4\\
$s5$ & 0.6 & 7.24 & 0 & 0 & 4\\
\hline
$s6$ & 0.3 & 7.24 & 1100 & 1.44 & 4\\
$s7$ & 0.3 & 7.24 & 1100 & 2.16 & 4\\
$s8$ & 0.3 & 7.24 & 1750 & 1.44 & 4\\
$s9$ & 0.3 & 7.24 & 1100 & 1.44 & 10\\
\hline
\end{tabular}
\label{tab1}
\end{center}
\end{table}

The reaction parameters chosen for the cases considered mimic the combustion
of a typical hydrocarbon in air at low to moderate values of Reynolds number.
For all reacting cases presented most of the reaction 
occurs between $2<St<8$ and the reaction rate peaks between $4<St<6$. 

\section{Results}

\begin{figure}[h]
\vskip 1.5cm
\centerline{\psfig{file=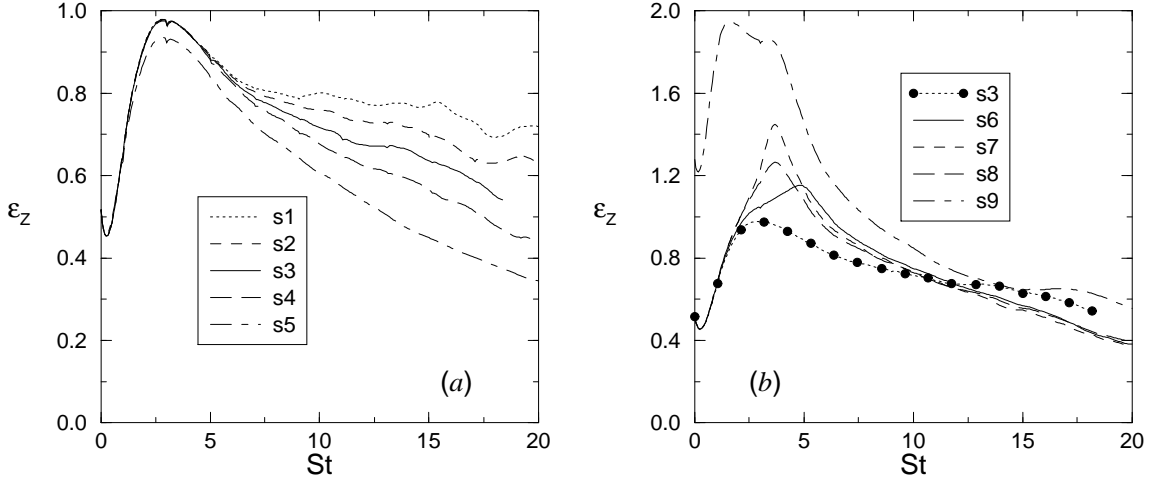,width=15cm}}
\caption{a) Mach number and b) heat release influence on the evolution of the
mixture fraction dissipation rate.}
\label{f1}
\end{figure}

In the absence of a mean scalar gradient, the transport equation for the 
mixture fraction variance, $\widetilde{Z''Z''}$, does not have a production 
term and $\widetilde{Z''Z''}$ decays continuously. Furthermore, if the 
transport equation for the mixture fraction variance is normalized by 
$\widetilde{Z''Z''}$ then the right hand side is proportional to the mixture 
fraction dissipation rate, 
$\epsilon_Z=\frac{1}{\widetilde{Z''Z''}}<\frac{\mu}{Re Sc}\nabla Z''\cdot
\nabla Z''>$ which is a key quantity in the modeling of both passive and 
reactive turbulent scalar fields. For all cases considered $\epsilon_Z$ 
increases initially as the turbulence breaks down the non-premixed ``blobs'' 
into smaller scalar structures, then it reaches a maximum and starts
to decay. For the isotropic turbulence cases our results indicate a slight 
dependence of $\epsilon_Z$ on $M_{t_0}$. However, for the shear flow 
simulations, figure \ref{f1}(a)  shows that the values of $\epsilon_Z$ 
decrease as $M_{t_0}$ increases. Since for the nonreacting cases the variation
 of the viscosity is small, the decrease in $\epsilon_Z$ with Mach number can 
be associated with the decrease in the normalized mixture fraction gradient 
variance, $\xi=<\nabla Z''\cdot\nabla Z''>/\widetilde{Z''Z''}$.
This is in agreement with the results obtained for
forced isotropic turbulence by Cai, O'Brien \& Ladeinde (1998). 
In the presence of heat release, $\epsilon_Z$ increases its magnitude during 
the time when the reaction is significant (figure \ref{f1}b). A further 
increase in the values of $\epsilon_Z$ is obtained  by increasing $Ce$ or 
$Da$. However, for the reacting cases,
our results indicate that the values of $\xi$ decrease due to the reaction. 
Therefore, the increase in $\epsilon_Z$ can be associated with the significant 
enhancement of the molecular transport properties due to the heat of reaction.
By increasing the value of $k_{0_s}$ (case $s9$) the initial length scale of 
the scalar field becomes smaller, which results in higher values of 
$\epsilon_Z$ compared to other cases.

\begin{figure}[h]
\vskip 1.5cm
\centerline{\psfig{file=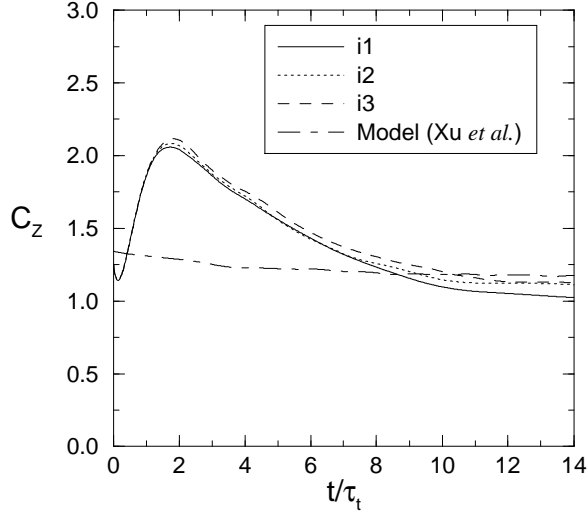,width=7.5cm}}
\caption{Time variation of $C_Z$ for the isotropic turbulence cases.}
\label{f2}
\end{figure}

The scalar dissipation rate is usually modeled by assuming that the ratio, 
$C_Z=\frac{2K/\epsilon}{1/\epsilon_Z}$, of the time scale of mechanical
dissipation to that of the mixture fraction variance dissipation is constant, 
although there is much evidence that it does not take a universal value 
(Livescu, Jaberi \& Madnia 2000). Figure \ref{f2} shows that for the isotropic
turbulence cases $C_Z$ varies in time initially and  seems to 
asymptote to a constant at long time. Similar with the results obtained for the
mixture fraction dissipation  rate, $C_Z$ is slightly affected by the change 
in $M_{t_0}$ for the cases considered. For 
comparison, also shown are the corresponding values of $C_Z$ obtained using the
model of Xu, Antonia \& Rajagopalan (2000). The model assumes local isotropy, 
and the prediction plotted in figure \ref{f2} does not take into account the 
intermittency of the small scales. The values of $C_Z$ become close to 
the model prediction at late times.

\begin{figure}[h]
\vskip 5.5cm
\centerline{\psfig{file=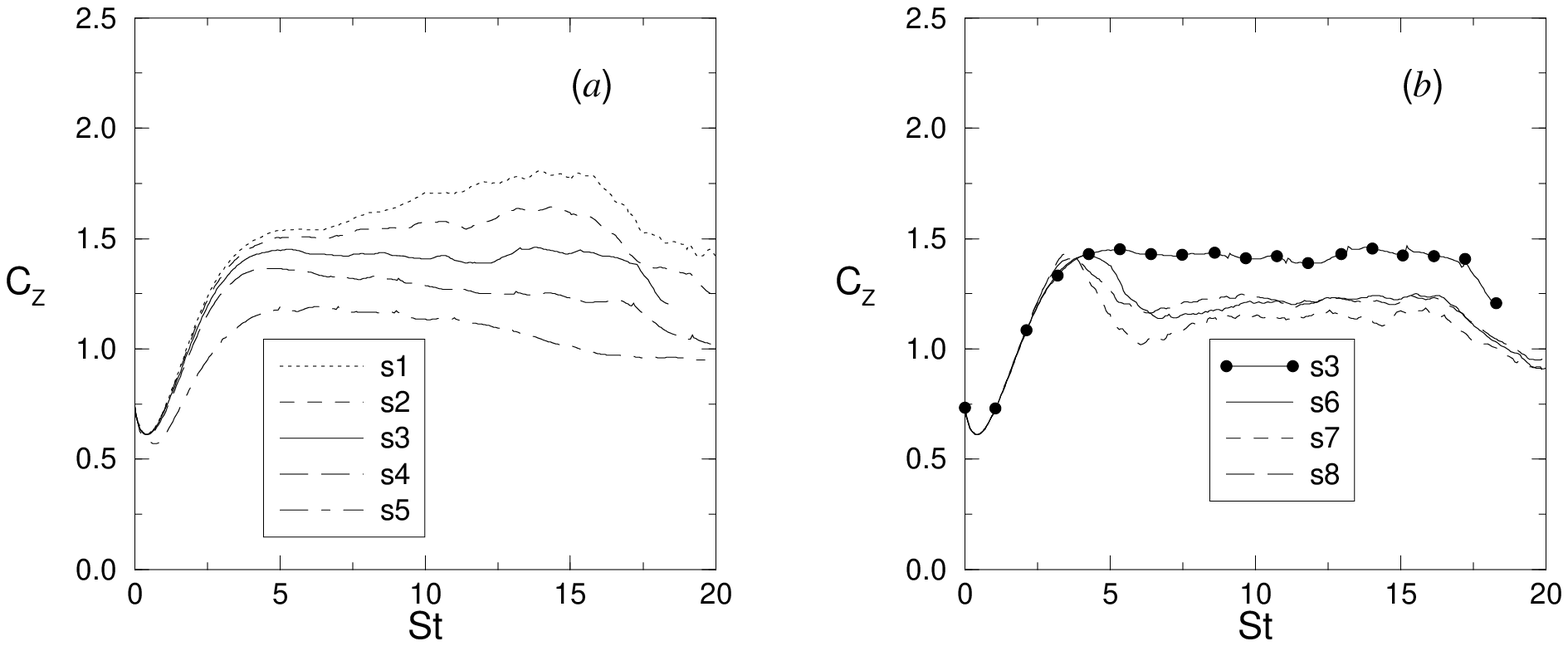,width=15cm}}
\caption{a) Mach number and b) heat release influence on the evolution of 
$C_Z$.}
\label{f3}
\end{figure}

Both $M_{t_0}$ and heat of reaction influence the values of $C_Z$ in the shear
flow simulations (figure \ref{f3}). For the nonreacting cases, $C_Z$ decreases 
as $M_{t_0}$ increases (figure \ref{f3}a). At early times this is due to an 
amplification of the viscous dissipation rate, $\epsilon/K$, and a 
decrease in $\epsilon_Z$. At later times $\epsilon/K$ has lower values at 
higher $M_{t_0}$, which explains the change in the slope of $C_Z$ observed in 
figure \ref{f3}(a). For the reacting cases the viscous dissipation rate 
increases 
significantly during the time when reaction is important. This increase is 
larger than the increase in $\epsilon_Z$, and $C_Z$ becomes less 
than in the nonreacting case (figure \ref{f3}b).

It is explained above that the normalized mixture fraction gradient, decreases
at higher values of $M_{t_0}$ or in the presence of heat release. This behavior
is further examined by considering the transport equation for the
normalized variance of the mixture fraction gradient:

\begin{equation}
 \frac{d\xi}{dt}=I+II+III+IV+V
\end{equation}

\newpage
\noindent
where
\begin{eqnarray}
I&=&-2<\nabla Z''\cdot\underline{\underline{S}}\cdot\nabla Z''>
  /\widetilde{Z''Z''}\nonumber \protect \\ 
II&=&-2<\nabla Z''\cdot \underline{\underline{s^s}}
  \cdot \nabla Z''>/ \widetilde{Z''Z''}\nonumber \protect \\ 
III&=&-2<\nabla Z''\cdot \underline{\underline{s^d}}\cdot\nabla Z''>
  /\widetilde{Z''Z''}\nonumber \protect \\  
IV&=&<\Delta (\nabla Z'' \cdot \nabla Z'')>
  /\widetilde{Z''Z''}\nonumber \protect \\ 
V&=&-\left[\frac{2 \tilde{\mu}}{Re Sc<\rho>}\left<\left(
 \nabla^2 Z''\right)^2-\frac{<\nabla Z'' \cdot \nabla Z''>^2}
 {\widetilde{Z''Z''}}\right>\right]/\widetilde{Z''Z''}\nonumber
\label{eq1}
\end{eqnarray}

\noindent
and the terms involving fluctuations of the viscosity are neglected. Here
$s_{ij}^d=\frac{1}{2}\left(\frac{\partial u_i''^d}{\partial x_j}+
\frac{\partial u_j''^d}{\partial x_i}\right)$ and
$s_{ij}^s=\frac{1}{2}\left(\frac{\partial u_i''^s}{\partial x_j}+
\frac{\partial u_j''^s}{\partial x_i}\right)$ are the dilatational and
solenoidal strain rate tensors, respectively, and 
$\Delta=\frac{\partial u_i''}{\partial x_i}$ is the dilatation. 
The first two terms in 
equation \ref{eq1} are production terms, due to the mean shear and 
solenoidal strain rate, respectively, terms III and IV are explicit 
dilatational terms and the last term is the molecular dissipation. For both 
reacting and nonreacting 
cases considered the explicit dilatational terms are found to be much smaller
than the rest of the terms in equation \ref{eq1}, and will not be shown.

\begin{figure}[h]
\vskip 1.5cm
\centerline{\psfig{file=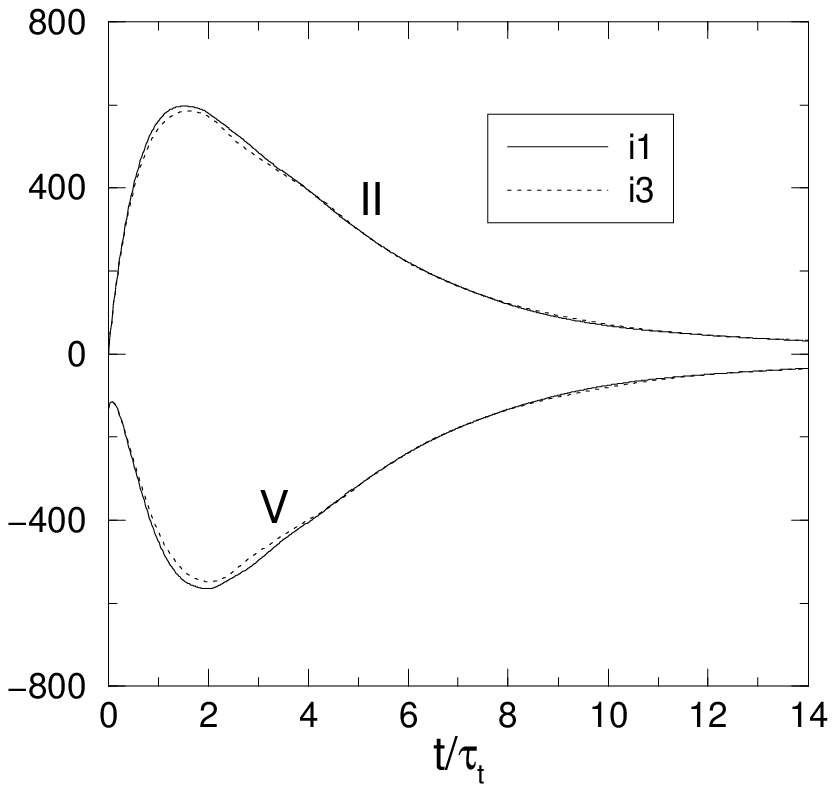,width=7.5cm}}
\caption{Important terms in equation \ref{eq1} for isotropic turbulence cases.}
\label{f4}
\end{figure}

Figure \ref{f4} shows that for isotropic turbulence the two important 
terms in equation \ref{eq1} (terms II and V), and therefore the rate of 
change of $\xi$, are slightly affected by the change in $M_{t_0}$.

\begin{figure}[h]
\vskip 1.5cm
\centerline{\psfig{file=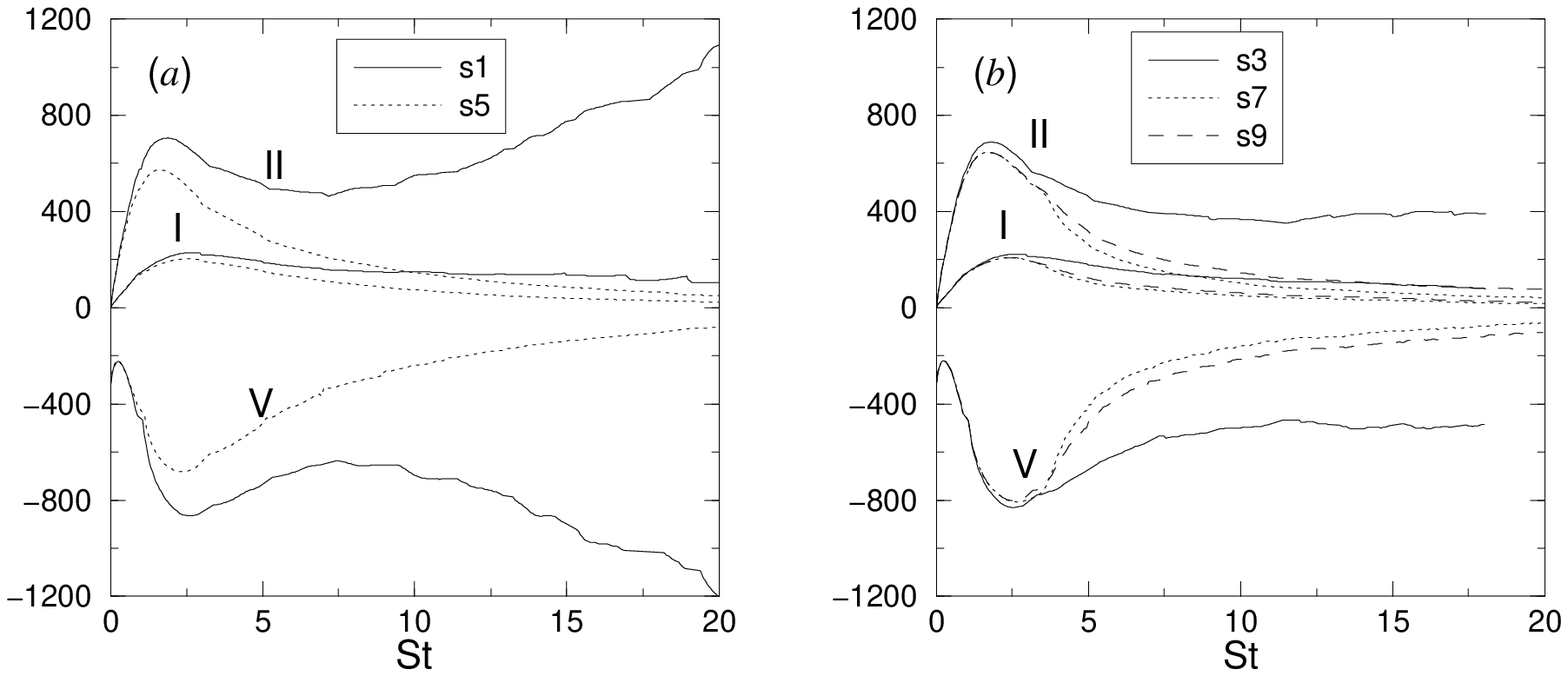,width=15cm}}
\caption{a) Mach number and b) heat release effect on the terms in equation 
\ref{eq1}.}
\label{f5}
\end{figure}

For the shear flow simulations, term I, production due to the mean shear is 
also important (figure \ref{f5}). However, for the value of $S_0^*$ considered 
for this study, this term is smaller than term II. Unlike the isotropic cases,
for the shear flow cases $M_{t_0}$ has a strong influence on the evolution of 
the terms in equation \ref{eq1} (figure \ref{f5}a). The production terms
and the dissipation term decrease their values significantly as $M_{t_0}$ 
increases. However, the decrease in the viscous dissipation term leads to an 
increase in the values of $\xi$. Therefore, the reduction in $\xi$ at higher 
$M_{t_0}$ is due to a decrease in the production terms, primarily due to the 
reduction of term II. Similarly, all terms in equation \ref{eq1} decrease 
their values for the reacting cases as compared to the nonreacting case 
(figure \ref{f5}b). Nevertheless, during the time when the reaction is 
significant, the reduction in the production terms (mainly term II) is 
responsible for the decrease in $\xi$.

The results presented suggest that the reduction in the values of $\xi$ 
(and hence $\epsilon_Z$ for the nonreacting cases) can be associated mainly to
a decrease in the production due to the solenoidal strain. This term is
dependent on the relative orientation between the mixture fraction gradient 
and the solenoidal strain rate tensor, the magnitude of the corresponding 
eigenvalues and the mixture fraction gradient variance. For compressible 
isotropic turbulence, Jaberi, Livescu \& Madnia (2000) showed that the scalar 
gradient tends to align with the most compressive eigenvector of the strain 
rate tensor, similar with the behavior observed earlier in incompressible 
turbulence (Ashurst {\it et al.} 1987). Our results indicate that 
this alignment is also obtained with the solenoidal part of the strain rate 
tensor and it does not change significantly with increasing $M_{t_0}$.

\begin{figure}[h]
\vskip 1.5cm
\centerline{\psfig{file=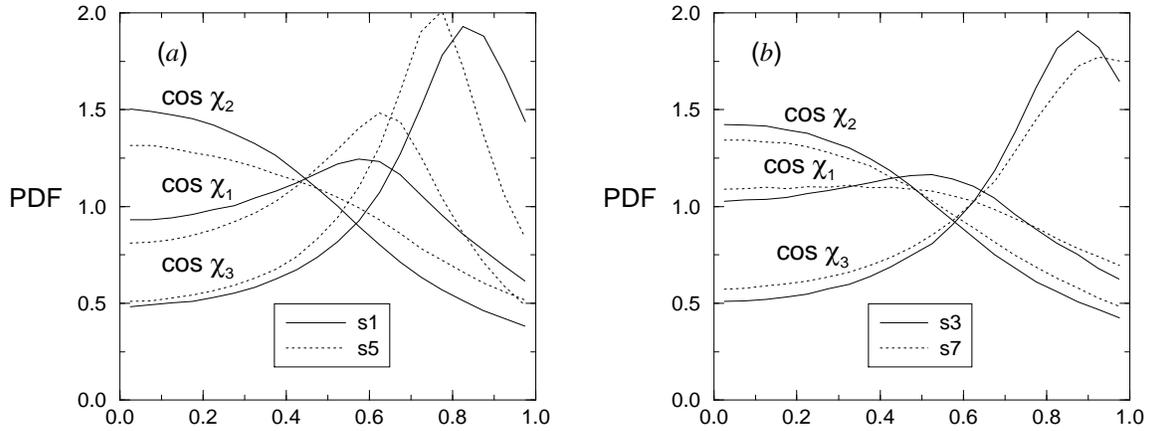,width=15cm}}
\caption{PDFs of the cosines of the angles between the eigenvectors of the 
solenoidal strain rate tensor and the mixture fraction gradient. a) Mach 
number effect ($St=8$) and b) heat release effect ($St=6$).}
\label{f7}
\end{figure}

However, for the shear flow simulations the PDF of the cosine of the angle 
between $\nabla Z''$ and the $\gamma$-eigenvector of 
$\underline{\underline{s^s}}$ peaks at a value different than 1, indicating a 
most probable distribution towards an angle different than zero 
(figure \ref{f7}). $\chi_1$, $\chi_2$, and $\chi_3$, are angles between 
$\nabla Z''$ and the $\alpha$-, $\beta$-, and $\gamma$-eigenvectors, 
respectively. These eigenvectors correspond to the eigenvalues labeled using 
the usual convention $\alpha >\beta >\gamma$. For a homogeneous flow, 
$<\alpha>+<\beta>+<\gamma>=0$. Since $<\alpha> >0$,  it has a negative 
contribution to the magnitude of the production term (term II in equation
\ref{eq1}), while $<\gamma>$ has a positive contribution. For all cases 
considered $<\beta> >0$ and it is small compared to $<\alpha>$ and $<\gamma>$.
As the initial turbulent Mach number increases, the peak of the PDF of 
$\cos \chi_3$ moves to smaller values, so the alignment worsens 
(figure \ref{f7}a), contributing to a decrease in the production term. 
Furthermore, the peak of the PDF of $\cos \chi_1$ tends to occur at higher 
values as $M_{t_0}$ increases, so that the alignment with the most dilatational
eigenvalue improves, further decreasing the production term. Moreover, all 
three eigenvalues of the solenoidal strain rate tensor decrease their magnitude
at higher $M_{t_0}$ and it can be shown that this effect also contributes to
the decrease in the production term.

For the reacting cases, near the time when the reaction rate peaks, the 
alignment between $\nabla Z''$ and the $\gamma$-eigenvector improves slightly
(the peak of the PDF of $\cos \chi_3$ moves to higher values), which
has a positive contribution to the production term (figure \ref{f7}b). 
Later, the alignment becomes close to that obtained for the nonreacting cases. 
However, the decrease in the magnitudes of the eigenvalues is more significant
and the production term has smaller values than in the nonreacting case.

\section{Conclusion}

DNS of compressible homogeneous turbulent shear flow and isotropic decaying
turbulence are performed under reacting (heat releasing) and nonreacting 
conditions to examine the role of compressibility and heat release on the
scalar mixing. For the reacting cases the chemical reaction is modeled as one 
step, irreversible, and Arrhenius type. The statistics
pertaining to the mixture fraction are extracted from the reacting scalars
and compared to those obtained for a passive scalar in the nonreacting
cases. 

For the range of $M_{t_0}$ examined, it is found that the mixture fraction 
dissipation 
rate is less sensitive to the changes in $M_{t_0}$ in isotropic turbulence 
than in shear flow. For the shear flow cases $\epsilon_Z$ is strongly 
affected by compressibility and it decreases as $M_{t_0}$ increases. Although 
$\epsilon_Z$ increases for the reacting cases during the time when the 
reaction is important, this is primarily due to the increase in the viscosity. 
Similar with the Mach number effect, the mixture fraction
gradient variance decreases for the reacting cases. The mechanical to scalar 
dissipation time scale ratio is also dependent on $M_{t_0}$ in shear flow, 
unlike the isotropic turbulence where the dependence was found to be weak. 

The transport equation for the normalized mixture fraction gradient variance 
was examined and was found that the explicit dilatational terms are much 
smaller than the other terms in the equation. For the shear flow simulations, 
the mixture fraction gradient variance decreases with Mach number primarily 
due to a misalignment between the mixture fraction gradient and the 
eigenvectors of the 
solenoidal strain rate tensor, and also a decrease in the magnitudes of the 
corresponding eigenvalues. However, for the reacting cases the normalized 
scalar gradient variance decreases mainly due to a reduction in the 
eigenvalues of the solenoidal strain rate tensor.

\section*{Acknowledgments}

This work is sponsored by the American Chemical Society under Grant 35064-AC9
and by the National Science Foundation under Grant CTS-9623178. 
Computational resources were provided by the National Center for Supercomputer
Applications at the University of Illinois at Urbana-Champaign, San Diego 
Supercomputer Center, and the Center for Computational Research at the State 
University of New York at Buffalo.

\end{document}